\documentclass[onecolumn,showpacs,preprintnumbers,amsmath,amssymb]{revtex4}

\usepackage{epsf}
\usepackage{graphicx}  
\usepackage{dcolumn}   
\usepackage{bm}        
\usepackage{subfigure}
\usepackage{multirow}

\def\aa{Astron. \& Astrophys.}

\def\apj{Astrophys. J.~}

\def\apjs{Astrophys. J. Suppl. Ser.~}

\def\jcap{J.~Cosmol.~Astropart.~Phys.~}

\def\plb{Phys. Lett. B~}
\def\prl{Phys. Rev. Lett.~}
\def\prd{Phys. Rev. D~}

\def\mnras{Mon. Not. R. Astron. Soc.~}

\begin{document}

\title{Galaxy clustering, CMB and supernova data constraints on $\phi$CDM model with massive neutrinos}

\author{Yun Chen}
 \email{chenyun@bao.ac.cn}
\affiliation{Key Laboratory for Computational Astrophysics, National Astronomical Observatories, Chinese Academy of Sciences, Beijing, 100012, P.~R.~China}

\author{Lixin Xu}
 \email{lxxu@dlut.edu.cn}
\affiliation{Institute of Theoretical Physics, School of Physics \& Optoelectronic Technology, Dalian
University of Technology, Dalian, 116024, P.~R.~China}
\affiliation{State Key Laboratory of Theoretical Physics, Institute of Theoretical Physics,
Chinese Academy of Sciences, Beijing, 100190, P.~R.~China}

\date{ \today}


\begin{abstract}

 We investigate a scalar field dark energy model (i.e., $\phi$CDM model) with massive neutrinos, where the scalar field possesses
an inverse power-law potential, i.e., $V(\phi)\propto {\phi}^{-\alpha}$ ($\alpha>0$). We find that the sum of neutrino
masses $\Sigma m_{\nu}$ has significant impacts on the CMB temperature power spectrum
and on the matter power spectrum. In addition, the parameter $\alpha$ also has slight impacts on the spectra. A joint sample, including CMB data from Planck 2013 and  WMAP9, galaxy clustering data from WiggleZ and BOSS DR11,
and JLA compilation of Type Ia supernova observations, is adopted to confine the parameters. Within the context of the $\phi$CDM model under consideration,
the joint sample determines the cosmological parameters to high precision: the angular size of the sound horizon at recombination,
the Thomson scattering optical depth due to reionization, the physical densities
of baryons and cold dark matter, and the scalar spectral index are estimated to be $\theta_* = (1.0415^{+0.0012}_{-0.0011})\times10^{-2}$,
$\tau = 0.0914^{+0.0266}_{-0.0242}$, $\Omega_b h^2 = 0.0222\pm 0.0005$, $\Omega_c h^2 = 0.1177 \pm 0.0036$, and $n_s = 0.9644^{+0.0118}_{-0.0119}$, respectively, at 95\% confidence level (CL).
It turns out that $\alpha <4.995$ at 95\% CL for the $\phi$CDM model. And yet, the $\Lambda$CDM scenario
 corresponding to $\alpha = 0$ is not ruled out at 95\% CL.
Moreover, we get $\Sigma m_{\nu}< 0.262$ eV
at 95\% CL for the $\phi$CDM model, while the corresponding one for the $\Lambda$CDM model is $\Sigma m_{\nu} < 0.293$ eV.
The allowed scale of $\Sigma m_\nu$ in the $\phi$CDM model is a bit smaller than that in the $\Lambda$CDM model.
It is consistent with the qualitative analysis, which reveals that the increases of $\alpha$ and $\Sigma m_\nu$ both
can result in the suppression of the matter power spectrum. As a consequence, when $\alpha$ is larger, in order to avoid suppressing the
matter power spectrum too much, the value of $\Sigma m_\nu$
should be smaller.

\end{abstract}


\maketitle

\section{INTRODUCTION}
\label{intro}

Neutrino is one of the important bonds linking nuclear physics, particle physics, astrophysics and cosmology \cite{Kamionkowski1999_Balantekin2013}.
In the Standard Model (SM) of particle physics, it is anticipated that there are three types, or ``flavors'', of  neutrinos: electron neutrino ($\nu_e$), muon neutrino ($\nu_{\mu}$) and tau neutrino ($\nu_{\tau}$), which are also dubbed as three normal/active neutrinos.
Besides that, neutrinos are assumed to be massless in the SM of particle physics \cite{Giacomelli2009}.

It was first predicted by  Bruno Pontecorvo in 1957 that if neutrinos are massive the neutrino flavor should be unstable, that is called
neutrino (flavor) oscillations \cite{Pontecorvo1957}. Briefly put, neutrino oscillation is a phenomenon that a neutrino
 produced in a definite flavor is observed in a different flavor after traveling some distances. In other words, neutrinos
 are able to oscillate among the three available flavors while they propagate through space. Nowadays there are compelling
 evidences for neutrino oscillations from a variety of experimental data on solar, atmospheric, reactor and accelerator neutrinos.
The discovery of neutrino oscillations implies that neutrinos have small but non-zero masses, with at least two species being
non-relativistic today. However, the present experimental results on neutrino oscillations only measure the difference
 of two squared masses, such as $\Delta m_{21}^2=m_2^2-m_1^2$ and $\Delta m_{32}^2=m_3^2-m_2^2$, but give no hint on their
  absolute mass scales.  $m_1$, $m_2$ and $m_3$ are the neutrino mass eigenstates. For example, the solar neutrino analysis supplemented by KamLAND produces an estimate
  of $\Delta m_{21}^2\sim 8\times 10^{-5}$eV$^2$ \cite{Araki2005_Abe2008}, and the measurement of atmospheric
  neutrino oscillation by Super-Kamiokande I indicates $\Delta m_{32}^2\sim 3\times 10^{-3}$eV$^2$ \cite{Ashie2005}.
  If it is the case of oscillations among three light neutrinos, only two of the three $\Delta m_{ij}^2$ are independent, as
$\Delta m_{21}^2+\Delta m_{32}^2+\Delta m_{13}^2=0$, where $\Delta m_{13}^2=m_1^2-m_3^2$. Recent reviews on progress in
both theoretical and experimental aspects of neutrino oscillations can be found in \cite{Maltoni2004_Fogli2006_Balantekin_Haxton2013}.

A variety of cosmological tests are sensitive to the absolute scale of neutrino mass, such as the cosmic microwave background (CMB)
radiation, galaxy surveys, and the Lyman-alpha forest \cite{Lesgourgues2006_2012}. In \cite{Roncarelli2015},
the effect of massive neutrinos on the Sunyaev--Zel'dovich and X-ray observables of galaxy clusters are investigated with
 a set of six very large cosmological simulations (8$h^{-3}$ Gpc$^3$ comoving volume). The analysis of current cosmological
 observations provides an upper bound on the total neutrino mass $\sum m_{\nu}$ (summed over the three neutrino families) of
 order 1 eV or less. However, the limits on $\sum m_{\nu}$ from cosmology are rather model dependent and vary strongly with
 the data combination adopted. For example, in the framework of one-parameter extensions to the base $\Lambda$CDM model the
Planck 2015 results \cite{Planck2015} give 95\% upper limits on the sum of neutrino masses, i.e., $\sum m_{\nu} < 0.23 $ eV
for a combination of Planck TT+lowP+lensing+ext, and $\sum m_{\nu} < 0.59 $ eV for Planck TT, TE, EE+lowP+lensing,
where ``TT'' denotes the combination of the TT likelihood at multipoles $l \geq 30$ and a low-\textit{l} temperature-only likelihood,
``TE'' denotes the likelihood at $l \geq 30$ using TE spectra, and ``EE''denotes the likelihood at $l \geq 30$ using EE spectra,``lowP''
denotes the low-\textit{l} Planck polarization data, ``lensing'' is the Planck lensing data, and ``ext'' represents the external data
including the baryon acoustic oscillations (BAO), Type Ia supernovae (SNe Ia), and $H_0$. In \cite{Geng2015}, the power law and exponential types of viable $f(R)$
theories along with massive neutrinos are studied. It shows that the allowed scales of  $\sum m_{\nu}$ in the viable $f(R)$ models are
greater than that in the $\Lambda$CDM model. The cases of fixing the effective number of neutrino
species as $N_\textrm{eff} = 3.046$ and treating $N_\textrm{eff}$ as a free parameters are both considered in \cite{Geng2015}.
The former corresponds to just consider the active neutrinos without the effect of dark radiation. The latter corresponds to
include the contribution of dark radiation (represented by $\Delta N_\textrm{eff} = N_\textrm{eff}-3.046$). For more details
on dark radiation, we refer the reader to \cite{Ackerman2009_Archidiacono2011}. The model of holographic dark energy
with massive neutrinos and/or dark radiation is investigated in \cite{Zhang2015}, but the computed results from this model are not
compared with those from the $\Lambda$CDM model. Actually, the $\Lambda$CDM model with massive neutrinos
is discussed broadly with constraints from various cosmological
observations \cite{Gratton2008_Riemer-Sorensen2014_Beutler2014_Palanque-Delabrouille2015}. The time evolving of neutrino mass
is also explored in the literature \cite{Fardon2004_Gu2004_Peccei2005_Zhao2007_Franca2009_Chitov2011}.
For further details on neutrino cosmology, the reader is referred to recent reviews
 such as \cite{Lesgourgues2006_2012, Dolgov2002_Hannestad2006_Hannestad2010_Wong2011}.

In this paper, we will discuss the constraints on the sum of neutrino masses $\sum m_{\nu}$ in the framework of $\phi$CDM model
by using a combination of the CMB data from Planck 2013 and WMAP9, the galaxy clustering data from WiggleZ and BOSS surveys,
 and the JLA compilation of  SNe Ia observations. The effect of dark radiation is not considered in this work, i.e., $N_\textrm{eff} = 3.046$.
 We also assume that one of the three active neutrinos is massive, and the other two are massless.
The $\phi$CDM model --- in which dark energy is modeled as a scalar
field $\phi$ with a gradually decreasing (in $\phi$) potential $V(\phi)$ --- is a simple dynamical model with
a slowly decreasing (in time) dark energy density. This model could resolve some of the puzzles of
the $\Lambda$CDM model \cite{Zlatev1999_ChenY2010}, such as the coincidence and fine-tuning problems. Here we focus on the scalar field with
an inverse  power-law potential $V(\phi)
\propto \phi^{-\alpha}$, where $\alpha$ is a nonnegative constant
\cite{Peebles&Ratra1988, Ratra&Peebles1988}. When $\alpha = 0$ the
$\phi$CDM model is reduced to the corresponding $\Lambda$CDM case. The $\phi$CDM model with this kind of $V(\phi)$
has been extensively investigated\cite{Avsajanishvili2014,Pavlov2014, phiCDM}, but
  without considering the massive neutrinos.

The rest of the paper is organized as follows. In Sec.\ {\ref{equations}} we present the background and perturbation evolutions of
the $\phi$CDM model with massive neutrinos. The impacts of $\sum m_{\nu}$ and $\alpha$ on the CMB temperature power spectrum and
 on the matter power spectrum are also discussed. Constraints from the cosmological data are derived in Sec.\
{\ref{Observation}}, and the results for $\phi$CDM model are compared with those for the $\Lambda$CDM model. We summarize our
 main conclusions in Sec.\ {\ref{summary}}.

%
%
\section{The $\phi$CDM model with massive neutrinos}
\label{equations}
\subsection{Background evolution of the $\phi$CDM model}
Quintessence as one of the popular scalar field dark energy models is a hypothetical form of dynamical dark energy to explain the late-time cosmic acceleration. Since quintessence
is described by the scalar field $\phi$, the corresponding dark energy model can also be
called as $\phi$CDM model. In what follows, we will use the terms ``quintessence'' and ``$\phi$CDM''
essentially interchangeably. We consider the self-interacting scalar field $\phi$ minimally coupled to gravity on cosmological scales. The action of this $\phi$CDM model is given by
\begin{equation}
\label{eq:totAction}
S=\int~\sqrt{-g}\left(-\frac{m^2_p}{16 \pi}R + \mathcal{L}_{\phi}+\mathcal{L}\right)d^4 x,
\end{equation}
where $g$ is the determinant of the metric $g_{\mu\nu}$, $R$ is the Ricci scalar,
$m_p=1/\sqrt{G}$ is the Planck mass with $G$ being the Newtonian constant of gravitation,
$\mathcal{L}$ is the Lagrangian density for matter and radiation, and $\mathcal{L}_{\phi}$
is the Lagrangian density for the field $\phi$, given by
\begin{equation}
\mathcal{L}_{\phi}=\frac{m^{2}_{p}}{16 \pi}\left[\frac{1}{2}g^{\mu \nu}\partial_{\mu}\phi\partial_{\nu}\phi-V(\phi)\right],
\label{eq:Lphi}
\end{equation}
where $V(\phi)$ is the field's potential.
In this work, we take a flat Friedmann-Lema\^{i}tre-Robertson-Walker (FLRW) metric for the background evolution, which is
described by
\begin{equation}
\label{eq:FLRWMetric}
ds^2=-dt^2+a^2\delta_{ij}dx^idx^j,
\end{equation}
where $x^i$ is the
comoving coordinate. $a(t)$ is the scale factor usually
normalized to unity now $a_0=a(z=0)=1$ and related to the redshift $z$ as $a/a_0=1/(1+z)$.
Throughout, the subscript ``0'' denotes the value of a quantity today.
By the variation of the action in
Eq.~(\ref{eq:totAction}) with respect to $\phi$, one can obtain the
Klein-Gordon equation (equation of motion) for the scalar
field
\begin{eqnarray}
\ddot\phi+3\left(\frac{\dot a}{a}\right)\dot \phi+\frac{d V}{d\phi}=0.
\label{eq:KGeq}
\end{eqnarray}
For the $\phi$CDM model, there are many kinds of $V(\phi)$ which can satisfy
the requirement of the late-time accelerating expansion of the universe~\cite{Samushia2009}.
In 1988, Peebles and Ratra~\cite{Peebles&Ratra1988} proposed a scalar field that is slowly rolling down
with a potential  $V(\phi)=\frac{1}{2}\kappa m^{2}_{p}\phi^{-\alpha}$ at a large $\phi$,
where $\kappa$ and $\alpha$ are nonnegative parameters. This inverse power-law potential
can not only lead to the late-time acceleration of the universe but also partially solve the cosmological
constant problems. The
larger value of $\alpha$ induces the stronger time dependence
of the scalar field energy density $\rho_{\phi}$. When $\alpha = 0$, this $\phi$CDM model is reduced to
the $\Lambda$CDM case. What is more, the parameter $\kappa$ depends on $\alpha$ (see \cite{FarooqPhDthesis, Avsajanishvili2014} for its dependence on
$\alpha$).

The Friedmann equation of the $\phi$CDM model with massive neutrinos can be written as
\begin{equation}
\label{eq:phiCDMFriedmann} H^2(z) =
\frac{8\pi}{3m_p^2}(\rho_b + \rho_c + \rho_{\phi} + \rho_{\gamma}+ \rho_{\nu}),
\end{equation}
where $\rho_b$, $\rho_c$, $\rho_\phi$, $\rho_{\gamma}$and $\rho_{\nu}$ denote the energy densities of baryons, cold dark matter (CDM), scalar field dark energy, photons and neutrinos,  and $H(z) \equiv \dot{a}/a$ is the Hubble parameter.
The energy density and pressure of the scalar field dark energy are given by
\begin{equation}
\label{eq:rhophi} \rho_{\phi} = \frac{m_p^2}{16\pi} (\dot{\phi}^2/2 +
V(\phi)),
\end{equation}
and
\begin{equation}
\label{eq:Pressurephi} P_{\phi} = \frac{m_p^2}{16\pi} (\dot{\phi}^2/2 -
V(\phi)).
\end{equation}
Then, one can work out the equation of state (EoS) of the field $\phi$,
\begin{equation}
\label{eq:EoSphi}
\omega_{\phi}\equiv P_{\phi}/\rho_{\phi}=\frac{\dot{\phi}^2-2V(\phi)}{\dot{\phi}^2+2V(\phi)},
\end{equation}
which is clearly bounded in the range $-1<\omega_{\phi}<1$ and usually non-constant. One can see that if the scalar field $\phi$ rolls slowly enough such that the kinetic energy density is much less than the potential energy density, i.e. $\dot{\phi}^2 \ll V(\phi)$, the pressure $P_{\phi}$ of the field will become negative with $\omega_{\phi}\rightarrow -1$.

Based on Eqs.~(\ref{eq:KGeq}) and (\ref{eq:phiCDMFriedmann}), along with the initial conditions described in
Refs.~\cite{Peebles&Ratra1988,Avsajanishvili2014}, one can numerically compute the Hubble parameter $H(z)$.
We also introduce the dimensionless density parameter for each component as, $\Omega_X = \rho_X/\rho_{cr}$, where the index ``X''  denotes the individual components,
such as radiation (``r''), neutrino (``$\nu$'') and matter (``m''). The critical energy density is expressed as $\rho_{cr} = 3H^2m_p^2/(8\pi)$. $\Omega_m$ is the energy density of matter
including both baryons and CDM. $\Omega_{\nu}$ is the total neutrino energy density
which scales as $\propto a^{-4}$  at early times, and thereafter evolves as $\propto a^{-3}$ after the non-relativistic transition.
One can see that the massive neutrinos behave like the radiation at early times and like the matter later.

\subsection{Cosmological perturbation of the $\phi$CDM model}
Let us consider perturbations of the flat FLRW metric in the Newtonian Gauge \cite{Ma_Bertschinger1995}.
In this gauge, the linear perturbed metric is given by
\begin{equation}
\label{eq:PerturbedMetric}
ds^2 =a^2(\eta)\left[-(1+2\Psi)d\eta^2 + (1+2\Phi)\delta_{ij}dx^idx^j\right],
\end{equation}
where the scalar perturbations are dominant over vector or tensor perturbations, and $\eta = \int a^{-1} dt$ is the conformal time. The Newtonian force $\Psi$ gives rise to
the dynamics of the perturbed fluids, while the curvature perturbation $\Phi$ measures the local energy density fluctuations.
The linear perturbation theory is a good tool both for describing
the early universe at any scales, and the recent universe on the largest
scales.

 It has been shown in \cite{Mainini2008_Wang_Fan2012} that for self-interacting scalar field dark energy models it is phenomenologically sufficient to regard the dark energy component as a perfect fluid. We treat each component in the universe as perfect fluid, including the baryon, CDM, photon, neutrino and scalar field dark energy.
In the perfect fluid approach, the perturbed Einstein equations lead to the following Eqs.~(\ref{eq:ContiEq}) - (\ref{eq:Psi_Phi}) in the \emph{Fourier space}:
\begin{equation}
\label{eq:ContiEq}
\delta_X' + 3 Ha(c_{s,X}^2 - \omega_X)\delta_X,
 = -(1+\omega_X)(\theta_X + 3\Phi ')
\end{equation}
\begin{equation}
\label{eq:EulerEq}
\theta_X' + \left[Ha(1-3\omega_X)+\frac{\omega_X'}{1+\omega_X} \right]\theta_X = k^2 \left(\frac{c_{s,X}^2}{1+\omega_X}\delta_X+\Psi + \sigma_X\right),
\end{equation}
\begin{equation}
\label{eq:PoissonEq}
k^2\Phi = 4\pi Ga^2\rho_i\left[ \delta_X + 3Ha(\omega_X+1)\theta_X/k^2\right],
\end{equation}
and
\begin{equation}
\label{eq:Psi_Phi}
\Psi = -\Phi.
\end{equation}
The great advantage of linear theory is to obtain independent equations
of evolution for each Fourier mode.
All of the perturbed quantities $(\delta_X, \theta_X, \Psi, \Phi, etc.)$ are functions of space $x$ and time $t$,
where $X$ denotes each perfect fluid composing the universe. In the linear perturbations, the anisotropic stress $\sigma_X$
is negligible for the perfect fluids. Note that a prime  represents a derivative with respect to the conformal time $\eta$.
The spatial variation of density fluctuations is expressed by the density contrast
 $\delta_X \equiv \delta \rho_X/\bar{\rho}_X =  (\rho_X-\bar{\rho}_X)/\bar{\rho}_X$, and $\bar{\rho}_X$
 is the background energy density of component $X$. In the approximation of negligible irrotational flow,
 the divergence of the peculiar velocity $v_X$, $\theta_X = \nabla \cdot \vec{v}_X$
can be used to describe the fluid motion. In the \emph{Fourier space},
we have $\theta_X \equiv i\vec{k} \cdot \vec{v}_X$. While $\omega_X \equiv \bar{P}_X/\bar{\rho}_X$ is
the equation of state of each component, and $c_{s,X}^2 \equiv \delta P_X/\delta \rho_X$ represents the sound
velocity. Eq. (\ref{eq:ContiEq}) is called as the (perturbed) \emph{continuity equation}, that states the conservation
 of local density. Eq. (\ref{eq:EulerEq}) is called as the \emph{Euler equation}, that represents the conservation of
 local energy momentum, and describes dynamics of perturbed fluids originated by the Newtonian force $\Psi$.
 The curvature perturbation $\Phi$ is constrained to the local inhomogeneity via
 the \emph{Poisson equation} Eq. (\ref{eq:PoissonEq}). We can get Eq. (\ref{eq:Psi_Phi}) under the assumption that the perturbed fluid remains a perfect fluid.
 These equations (\ref{eq:ContiEq}) - (\ref{eq:Psi_Phi}) completely determine the dynamical evolution of
 large scale structure (LSS) of the universe, within a given expansion history $H$.

\subsection{Matter power spectrum and CMB power spectrum in the $\phi$CDM model}
In the framework of $\phi$CDM model, we qualitatively investigate the impacts of parameters $\alpha$ and $\Sigma m_{\nu}$
on the matter power spectrum and on the CMB power spectrum. The analyses are performed with the CAMB Boltzmann code \cite{Lewis2000}.

\begin{figure}[t]
\centering $\begin{array}{cc}
\includegraphics[width=0.4\textwidth]{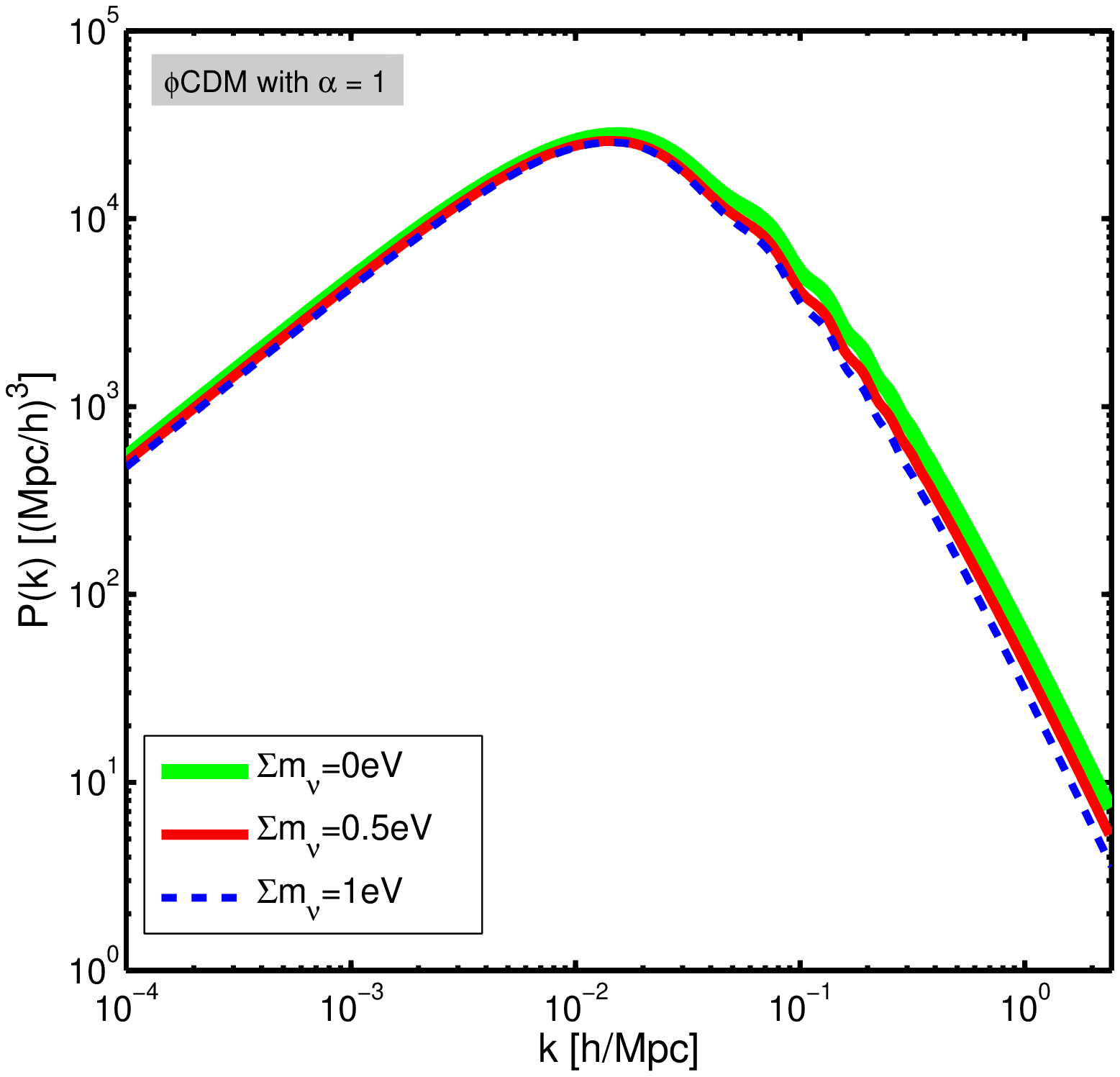}
\includegraphics[width=0.4\textwidth]{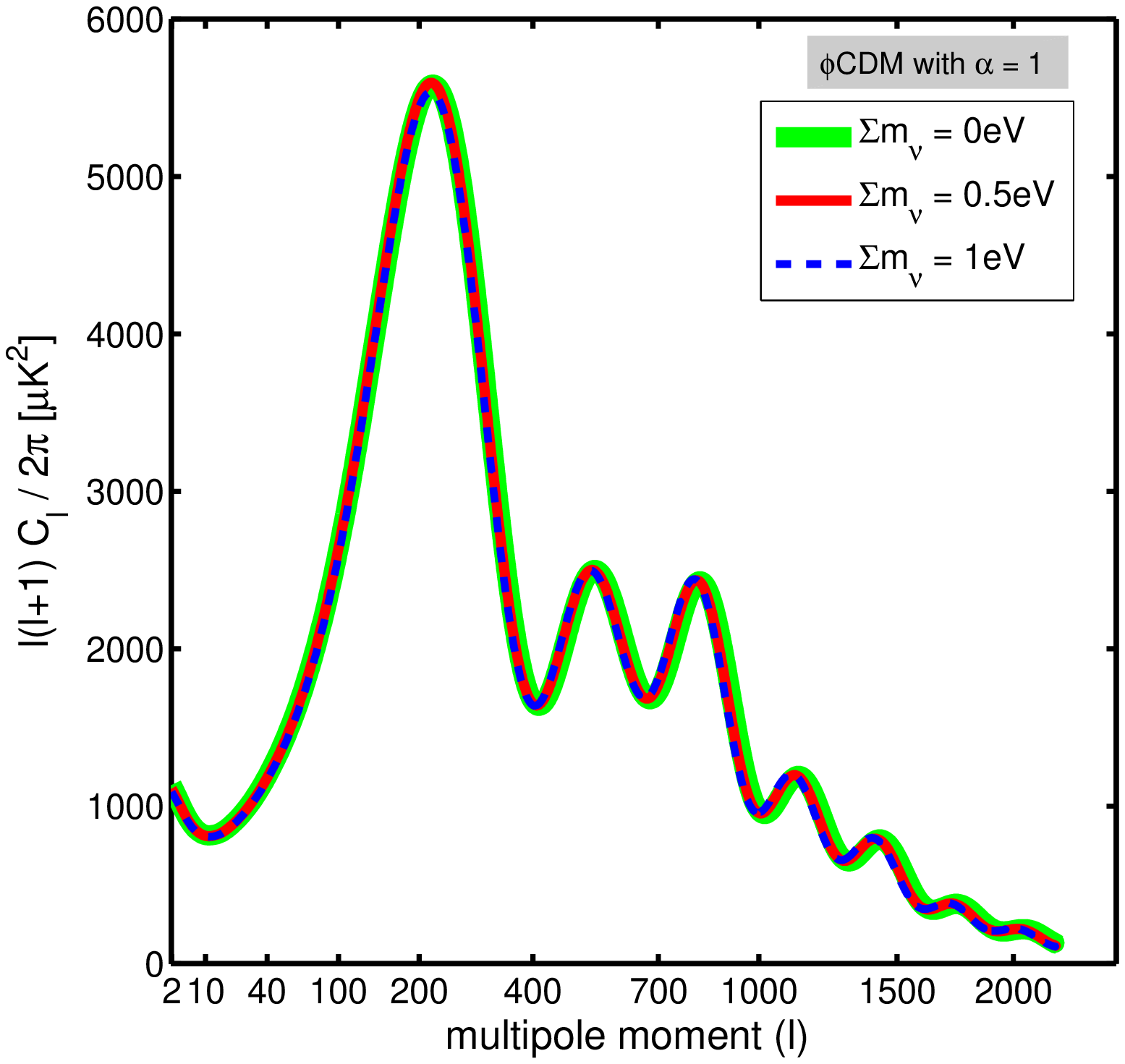} \\
\includegraphics[width=0.4\textwidth]{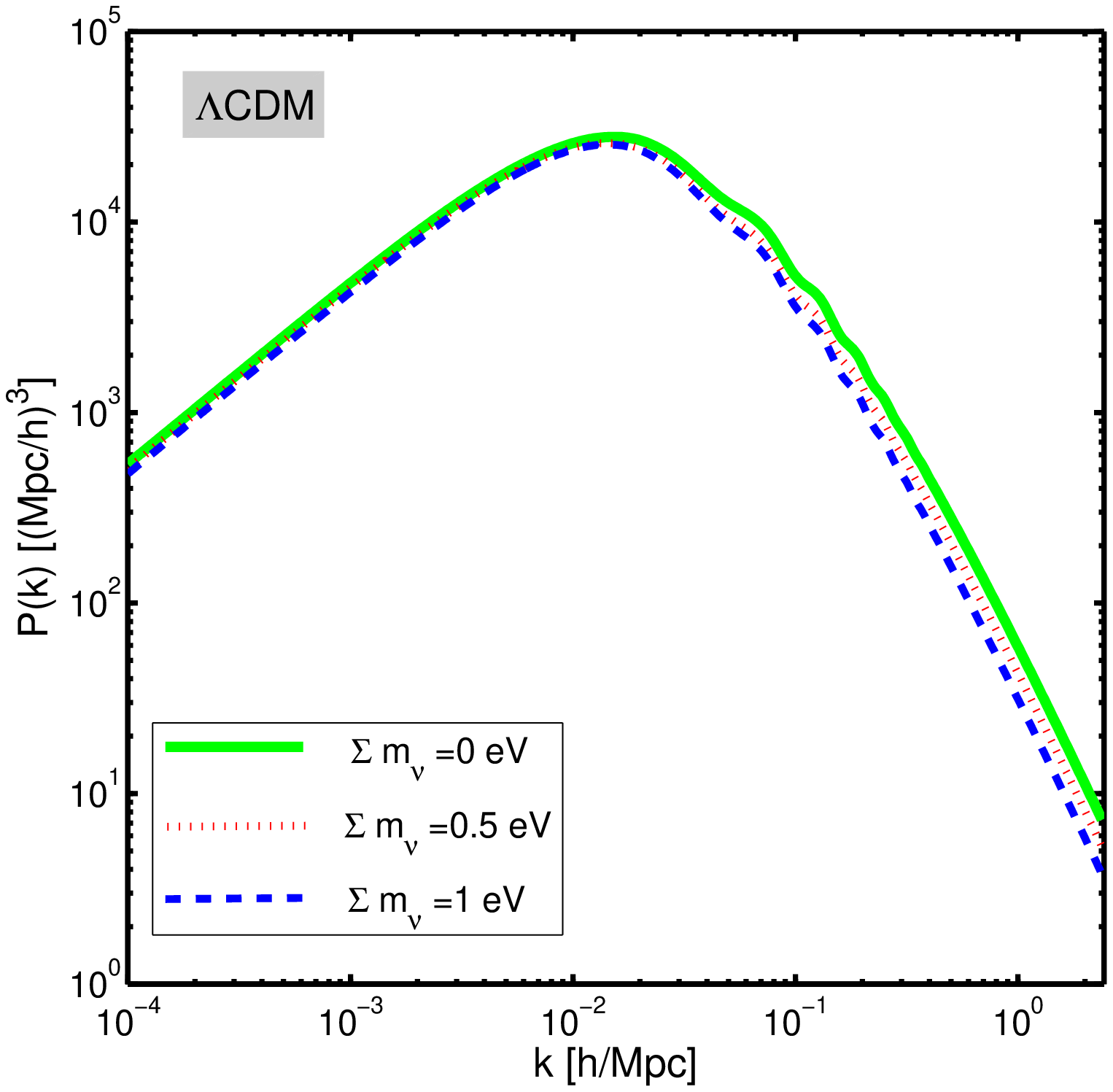}
\includegraphics[width=0.4\textwidth]{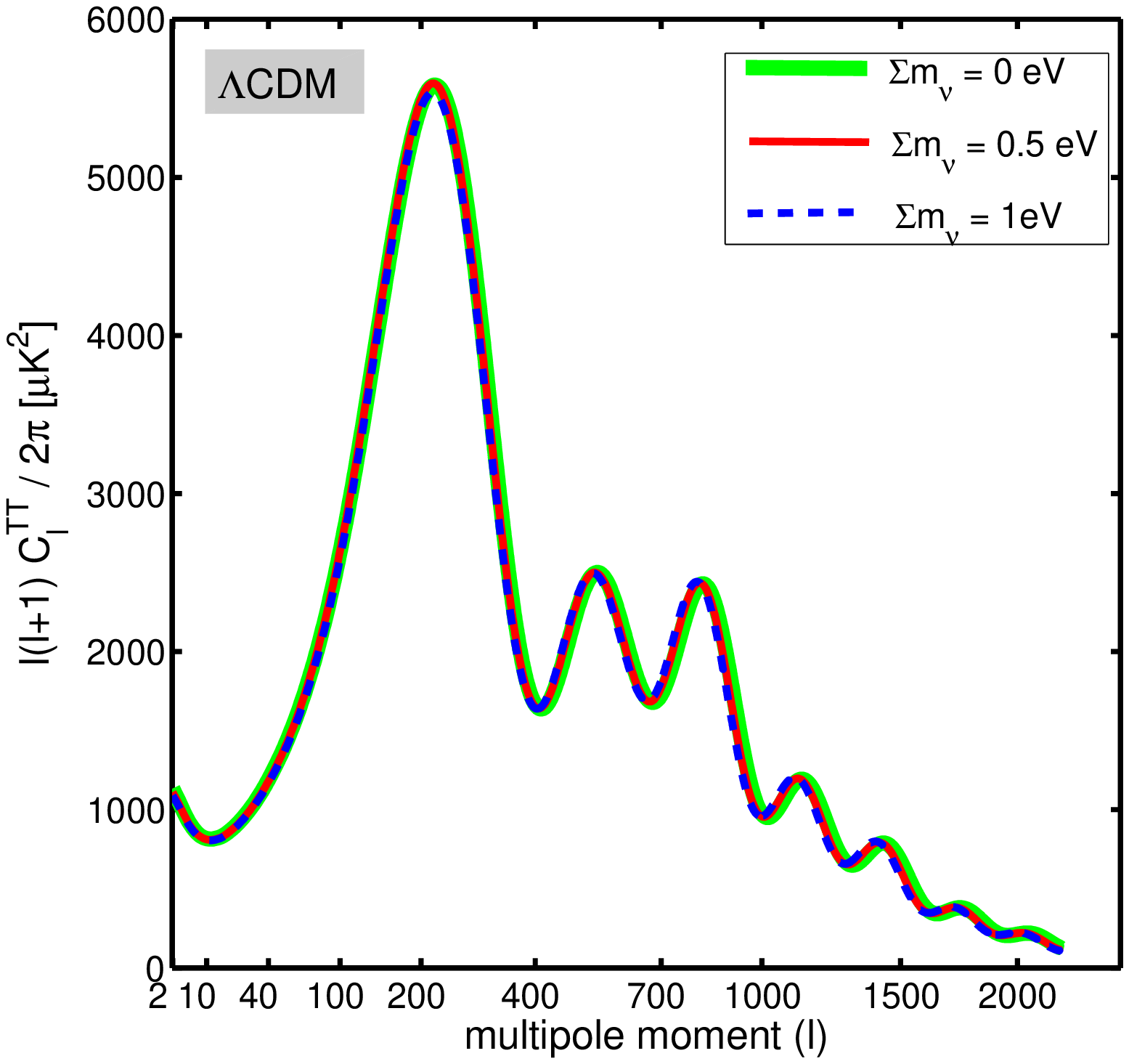}
\end{array}$
 \caption{Impacts of the sum of neutrino masses $\Sigma m_{\nu}$ on the matter power spectrum $P(k)$ and
 on the CMB temperature power spectrum $C_l^{TT}$ in the $\phi$CDM (upper panels) and $\Lambda$CDM (lower panels) models.
 $\Sigma m_{\nu}$ is varied, and other parameters are kept fixed.}
\label{fig:Pk_TT_mnu}
\end{figure}

 Neutrinos rarely  interact with matter after thermal decoupling, so they are treated as free streaming particles.
 Massive neutrinos are the only particles that present the transition from radiation to
matter. Before the non-relativistic transition the neutrinos behave like radiation.
Thus, when the neutrino mass $\Sigma m_{\nu}$ increases, the time of radiation/matter equality is postponed gradually,
and $a_{eq}$ increases. The value of  $\Sigma m_{\nu}$ can affect the matter power spectrum and the CMB power spectrum
 mainly resulting from a change in the time of equality, that provides a potential way to constrain it through CMB and LSS observations \cite{Lesgourgues2006_2012,Zhang2015,Bashinsky2004}. In Fig. \ref{fig:Pk_TT_mnu}, we show the impacts of neutrino mass $\Sigma m_{\nu}$ on the matter power spectrum $P(k)$ and the CMB temperature anisotropy spectrum $C_l^{TT}$. The upper panels show the cases for $\phi$CDM model with varying values of $\Sigma m_{\nu}$, where $\alpha$ is fixed as $\alpha = 1$, and other parameters are fixed based on the recent Planck results \cite{Planck2015}. For comparison, we also display the cases for $\Lambda$CDM model in the lower panels. For both $\phi$CDM and  $\Lambda$CDM models, the matter power spectrum is gradually suppressed with the increase of $\Sigma m_{\nu}$, however, the effect is more significant on small scales than that on large scales. One possible reason is that the neutrino perturbations do not contribute to gravitational clustering on scales smaller than the free-streaming scale, while on the very large scales neutrino perturbations are never affected by free streaming, and they become indistinguishable from CDM perturbations in the non-relativistic regime \cite{Lesgourgues2006_2012}. The CMB temperature anisotropy spectrum $C_l^{TT}$ is insensitive to the variation of $\Sigma m_{\nu}$ in both $\Lambda$CDM and $\phi$CDM models.

\begin{figure}[t]
\centering $\begin{array}{cc}
\includegraphics[width=0.45\textwidth]{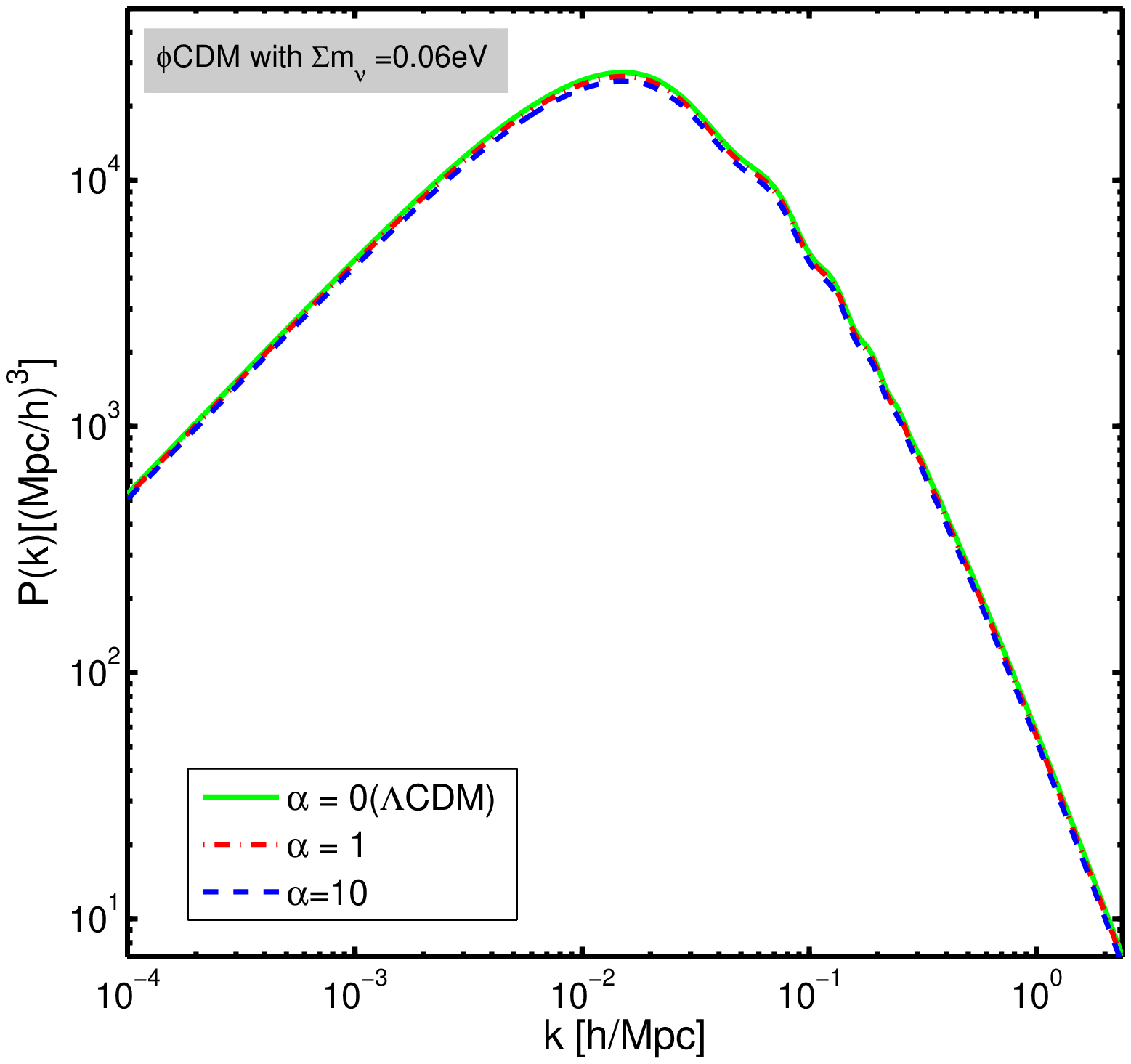}
\includegraphics[width=0.45\textwidth]{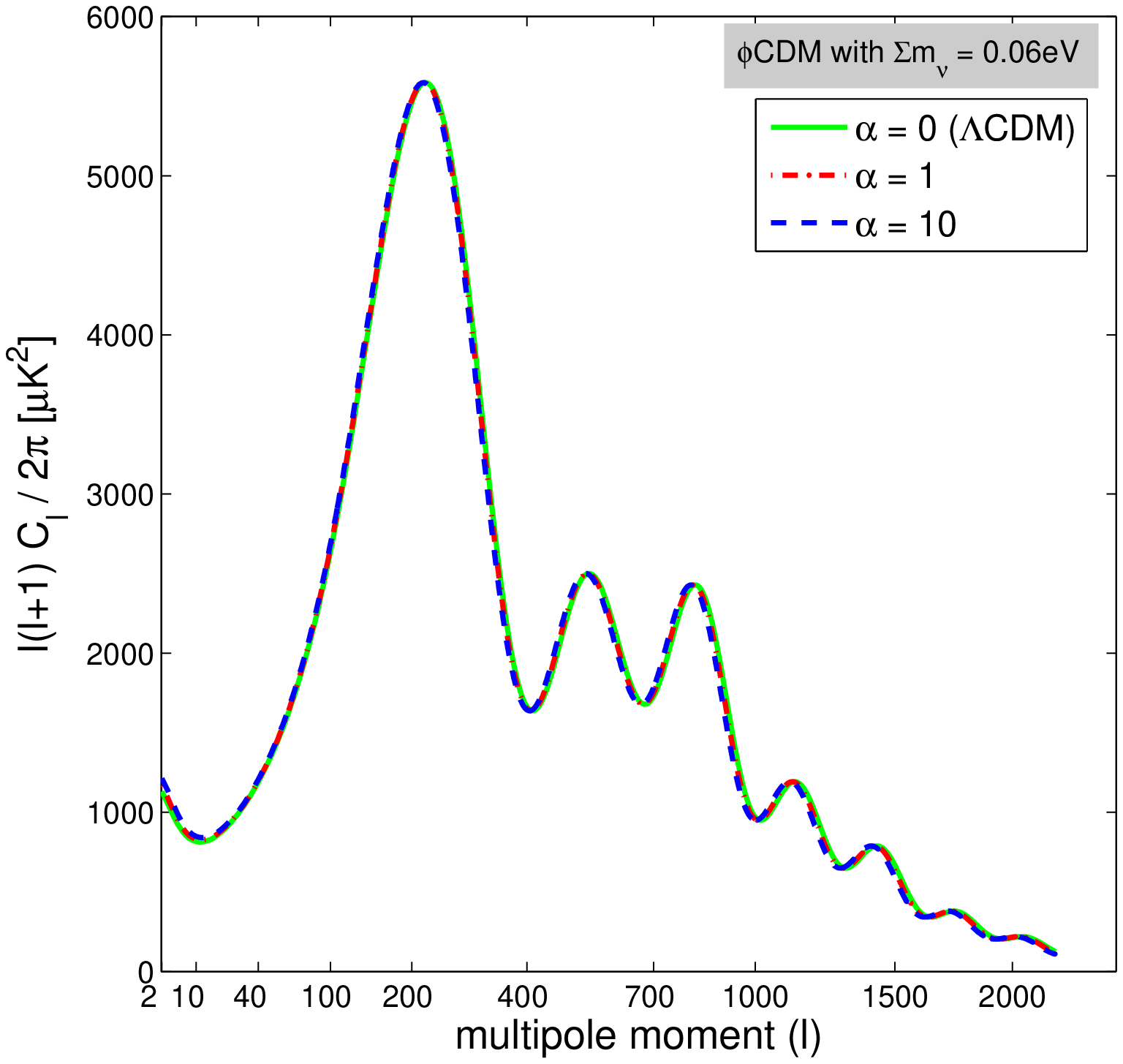}
\end{array}$
 \caption{Impacts of the parameter $\alpha$ on the matter power spectrum $P(k)$ and
 on the CMB temperature power spectrum $C_l^{TT}$ in the framework of $\phi$CDM model.
  $\alpha$ is varied, and other parameters are kept fixed.}
\label{fig:phiCDM_alpha_Pk_TT}
\end{figure}

The parameter $\alpha$ indicates the dynamics of dark energy, and then it can affect the expansion history of
the universe and the redshift of matter/dark energy equality. When $\alpha$ increases,
the expansion of the universe occurs more rapidly, and the epoch of dark energy domination begins
earlier \cite{Avsajanishvili2014}. For these reasons, the variation of $\alpha$ can have signatures in
the CMB map and the matter clustering. The impacts of $\alpha$ on $P(k)$ and $C_l^{TT}$
are presented in Fig. \ref{fig:phiCDM_alpha_Pk_TT}. We choose $\alpha =$ 0, 1, and 10 as examples,
where $\alpha = 0$ corresponds to the $\Lambda$CDM scenario. The values of other parameters are kept fixed.
We find that $P(k)$ is slightly suppressed with the increase of $\alpha$, and the effect is a bit significant
 on large scales than that on small scales. The CMB temperature anisotropy spectrum $C_l^{TT}$ is a little sensitive to the variation of $\alpha$ on the low-l tail,
  which may arise from the late Integrated Sachs-Wolfe (ISW) effect.
Anyhow, CMB and LSS observations are efficient to distinguish between $\Lambda$CDM and $\phi$CDM models.


\section{Observational constraints}
\label{Observation}

The observational data sets used to constrain the cosmological parameters are described as follows, including
 the galaxy clustering, CMB and SNe Ia measurements.

\subsection{Cosmological data sets}
\subsubsection{Galaxy clustering measurements}
Galaxy clustering distilled from the galaxy redshift survey is powerful as cosmological probe \cite{WangY2012}, that can allow us
to measure the cosmic expansion history through the measurement of BAO, and the growth history of
 cosmic large scale structure through measurements of redshift-space distortions.
The length scale of BAO, the comoving sound horizon at the baryon drag epoch $r_s(z_d)$, can be applied as a cosmological ruler and
accurately calibrated by observations of the CMB radiation. The position of the BAO peak in the
angle-averaged galaxy clustering pattern is usually quantified in term of the volume averaged distance \cite{Eisenstein2005}
\begin{equation}
\label{eq:DV}
D_V=\left[(1+z)^2D_A^2cz/H(z)\right]^{1/3}.
\end{equation}
It is common to report the BAO distance measurements as combinations of the angular
  diameter distance, $D_A(z)$, and the Hubble parameter, $H(z)$, such as
\begin{equation}
\label{eq:A_bao}
A(z) \equiv \frac{H_0 \sqrt{\Omega_{m}^0} D_V(z)}{c z},
\end{equation}
or
\begin{equation}
\label{eq:dz_bao}
d_z \equiv D_V(z)/r_s(z_d).
\end{equation}
The redshifts of galaxies include indistinguishable contributions from both the Hubble recession
 and the peculiar velocity of the galaxies themselves, so that there are errors in the distances we assign to galaxies.
  The differences between
 the redshift-inferred distances and true distances are known as redshift-space distortions (RSD) \cite{delaTorre2012}.
 In another word, the RSD are introduced in the observed clustering pattern by galaxy peculiar motions.
 As a consequence, the correlation function and the power spectrum measured in the redshift space are different
 from those in the real space, which have to be corrected to be expressed in real space. Because the effects of RSD
 couple the density and velocity fields, the RSD signals within the correlation function are difficult to model.
 On different scales, peculiar motions produce different types of distortions to the power spectrum.
 On small scales, i.e., in the cluster cores, the peculiar velocities of galaxies are almost randomly oriented,
that cause the structures to appear elongated along the line of sight (LOS) when viewed
in redshift space (i.e., the so called ``finger of God'' effect) \cite{Jackson1972},
leading to a damping of the clustering. On large scales, because of gravitational growth,
the galaxies tend to fall towards high-density regions, and flow away from low-density regions,
such that the galaxy clustering in redshift space is enhanced in the LOS direction compared to the transverse
direction \cite{Kaiser1987}. The RSD effect on large scales can be described
by linear theory \cite{Kaiser1987, Hamilton1992_1998}, while the ``finger of God''
effect is a non-linear phenomenon. On large scales where the gravitational growth is linear,
measuring the relative clustering in both LOS and transverse directions leads to measurements
of the parameter combination $f(z_{\textrm{eff}})\sigma_8(z_{\textrm{eff}})$, where $z_{\textrm{eff}}$
is the effective redshift.  $f$ is the growth rate of cosmic structure, which is associated with
the evolution of matter density perturbations $\delta_m$ via the relation $f \equiv d\ln \delta_m/d\ln a$.
In the linear regime, the linear growth rate can be expressed as  $f = d\ln D(z)/d\ln a$,
where $D(z)=\delta_m(z)/\delta_m(z=0)$ is the linear growth factor normalized such
that $D(z=0) = 1$. $\sigma_8(z)$ is the root-mean-square amplitude of the matter fluctuations in
spheres of 8$h^{-1}$ Mpc, and $\sigma_8(z=0)/\sigma_8(z)=D(z=0)/D(z)$. Thus, one can figure out
\begin{equation}
\label{eq:f_sigma8}
f(z)\sigma_8(z) = \sigma_8^0\frac{d D(z)}{d\ln a},
\end{equation}
where $\sigma_8^0 = \sigma_8(z=0)$.
In linear theory, the galaxy bias $b$ and the growth rate $f$ are degenerate with $\sigma_8$,
so the RSD measurements are better presented in terms of $b(z)\sigma_8(z)$ and $f(z)\sigma_8(z)$,
 rather than $f(z)$. Currently, the bias-independent parameter combination $f(z)\sigma_8(z)$ measured by RSD are widely used
 \cite{Pavlov2014, Blake2011, fsigma8}.

The Alcock-Paczynski (AP) test \cite{AP1979} is proved to be a significant link between BAO and RSD.
AP test states that if an astrophysical structure is spherically symmetric or isotropic, then it should
possess equal comoving transverse and radial sizes. An AP measurement is carried out by comparing the
observed transverse and radial dimensions of objects. While the AP test is equally valid for an isotropic
process such as the two-point statistics of galaxy clustering. The apparent anisotropy of the two-dimensional
correlation function of galaxies mainly arises from the geometry and expansion of the universe which should
be correctly embodied in the fiducial cosmological model, and the RSD effect which is supposed to
be marginalized by using appropriate RSD model (see \cite{delaTorre2012} for a recent review of RSD models).
According to the requirement of AP test, the signature of BAO should have identical comoving sizes (i.e., $r_s$)
in transverse and radial dimensions. The observed transverse dimension
is the angular projection $\Delta \theta = r_s/[(1+z)D_A(z)]$. The radial one is the redshift projection
$\Delta z = r_sH(z)/c$. The relative radial/transverse distortion depends on the value of
\begin{eqnarray}
\label{eq:Fz}
F(z)&\equiv& \Delta z/\Delta \theta\nonumber \\
    &=&(1+z)D_A(z)H(z)/c,
\end{eqnarray}
where $F(z)$ is dubbed as the AP distortion parameter.

By combining the BAO peak, AP test and RSD effect, one can report the galaxy clustering effectively as joint measurements of  $(A, F, f\sigma_8)$ or $(d_z, F, f\sigma_8)$.   These joint measurements are extremely good at helping to constrain basic cosmological parameters and distinguish between the dark energy models. By using large-scale structure measurements from the WiggleZ Dark Energy Survey \cite{Drinkwater2010}, Blake et al. (2012) \cite{Blake2012} have performed joint constraints of  $(A, F, f\sigma_8)$ in three overlapping redshift slices with effective redshifts $z_{\textrm{eff}} = (0.44, 0.6, 0.73)$. Utilizing these data, it is straightforward to put constraints on the model parameters by calculating the corresponding $\chi^2_{\textrm{WiggleZ}}$, given by
\begin{equation}
\label{eq:chi2_WiggleZ}
\chi^2_{\textrm{WiggleZ}} =(\vec{X}_{\textrm{obs}}-\vec{X}_{\textrm{th}}) \underline{C}^{-1} (\vec{X}_{\textrm{ob}s}-\vec{X}_{\textrm{th}})^T.
\end{equation}
The observational data vector is
\begin{equation}
\label{eq:X_obs}
\vec{X}_{\textrm{obs}} = [A_1, A_2, A_3, F_1, F_2, F_3, f\sigma_{8,1}, f\sigma_{8,2}, f\sigma_{8,3}],
\end{equation}
i.e., $\vec{X}_{\textrm{obs}}=[0.474, 0.442, 0.424, 0.482, 0.650, 0.865, 0.413, 0.390, 0.437]$ by using
the maximum likelihood estimations of $(A, F, f\sigma_8)$ listed in Table 1 of \cite{Blake2012}. The vector of theoretical values is
\begin{equation}
\label{eq:X_th}
\vec{X}_{\textrm{th}} = [A(z_1), A(z_2), A(z_3), F(z_1), F(z_2), F(z_3), f\sigma_8(z_1), f\sigma_8(z_2), f\sigma_8(z_3)],
\end{equation}
 where $[z_1, z_2, z_3] = [0.44, 0.6, 0.73]$, and the corresponding theoretical values of $(A, F, f\sigma_8)$ can
 be obtained with Eqs.(\ref{eq:A_bao}), (\ref{eq:Fz}) and (\ref{eq:f_sigma8}), respectively. $\underline{C}$ is
 a $9 \times 9$ covariance matrix between parameters and redshift slices, and the value of $10^3\underline{C}$ is
 listed in Table 2 of  \cite{Blake2012}, that is achieved by generating 400 lognormal realizations for each WiggleZ survey
 region and redshift slice with the methods described in \cite{Blake2011}. In the analysis of \cite{Blake2012}, the fitting formulae
provided by Jennings et al. (2011) \cite{Jennings2011} have been taken as the fiducial RSD model.
The effect of different choices of the RSD model is also considered in Section 3.4 of \cite{Blake2012}.
It turns out that the systematic error induced from modeling RSD is much lower than the statistical error in the measurement.

 Joint measurements of $(d_z, F, f\sigma_8)$ at an effective redshift of $z_{\textrm{eff}} = 0.57$ are
 provided in Samushia et al (2014) \cite{Samushia2014} by utilizing the observed anisotropic clustering
 of galaxies in the Baryon Oscillation Spectroscopic Survey (BOSS) Data Release 11 (DR11) CMASS sample \cite{Anderson2014}.
 We employ this data set in our analysis with the chi-squared statistic
 \begin{equation}
\label{eq:chi2_BOSS}
  \chi^2_{\textrm{BOSS}} = (\vec{Y}_{\textrm{obs}} - \vec{Y}_{\textrm{th}})  Cov^{-1} (\vec{Y}_{\textrm{obs}} - \vec{Y}_{\textrm{th}})^T.
\end{equation}
 The observational data vector is $\vec{Y}_{\textrm{obs}} = [d_z, F, f\sigma_8]$, i.e., $\vec{Y}_{\textrm{obs}} = [13.85, 0.6725, 0.4412]$
 by using the mean values presented in Eq. (30) of \cite{Samushia2014}.  The vector of theoretical values
 is $\vec{Y}_{\textrm{th}} = [d_z(z_{\textrm{eff}}), F(z_{\textrm{eff}}), f\sigma_8(z_{\textrm{eff}})]$,
 where the corresponding theoretical values of $(d_z, F, f\sigma_8)$ can be obtained with
Eqs.(\ref{eq:dz_bao}), (\ref{eq:Fz}) and (\ref{eq:f_sigma8}), respectively. The covariance matrix $Cov$ of measurements
is listed in Eq. (31) of \cite{Samushia2014}.  A suite of 600 PTHalo simulations are used to estimate
the covariance matrix (see \cite{Manera2013} for details of mock generation). In the analysis of \cite{Samushia2014},
the ``streaming model''-based approach developed in \cite{Reid2011} has been adopted to model the RSD signal,
that has been demonstrated to fit the monopole and quadrupole of the galaxy correlation function with better
than percent level precision to scales above $25h^{-1}$ Mpc, for galaxies with bias of $b \simeq 2$.

 The galaxy clustering (GC) measurements from WiggleZ and BOSS DR11 are both employed in this study. Thus, the corresponding
 chi-squared statistic is expressed as
 \begin{equation}
\label{eq:chi2_GC}
  \chi^2_{\textrm{GC}} = \chi^2_{\textrm{WiggleZ}} + \chi^2_{\textrm{BOSS}},
\end{equation}
where $\chi^2_{\textrm{WiggleZ}}$ and $\chi^2_{\textrm{BOSS}}$ are given by Eqs. (\ref{eq:chi2_WiggleZ}) and (\ref{eq:chi2_BOSS}),
respectively.

\subsubsection{CMB power spectrum measurements}

The CMB radiation deemed as the afterglow of the big bang can supply us with some information of the very early universe.
The observations of CMB provide another independent test for the existence of dark energy. The recent precise measurements
of the CMB radiation from Planck and Wilkinson Microwave Anisotropy Probe (WMAP) projects can efficiently improve the accuracy
 of constraining the cosmological parameters. Currently, the Planck 2015 results have come out \cite{Planck2015},
 but the Planck 2015 likelihoods are not yet available. Given this, we use the low multipoles ($2 \leq l \leq 49$)
 and high multipoles ( $50 \leq l \leq 2479$) temperature power spectrum likelihoods from Planck 2013 \cite{Planck2013PS_CP},
 together with the low multipoles ($l \leq 23$) polarization power spectrum likelihoods from nine-year WMAP (WMAP9) \cite{WMAP9FR_CP}.
To employ the previously mentioned CMB power spectrum data in the analysis, we compute the $\chi^2_{\textrm{CMB}}$ statistic
 \begin{equation}
\label{eq:chi2_CMB}
  \chi^2_{\textrm{CMB}} = \sum_{ll'}(C_l^{\textrm{obs}} - C_l^{\textrm{th}})\mathcal{M}^{-1}_{ll'}(C_{l'}^{\textrm{obs}} - C_{l'}^{\textrm{th}}),
\end{equation}
where $C_l^{\textrm{obs}}$ is the observational value of the related power spectrum,  $C_l^{\textrm{th}}$ is the corresponding
theoretical value in the framework of the cosmological model under consideration,
and $\mathcal{M}$ is the covariance matrix for the best-fit data spectrum.

\subsubsection{ Magnitude-redshift measurements of Type Ia supernovae}

The first direct evidence for the cosmic acceleration
came from SNe Ia observations, which
provide the measurement of the cosmic expansion history through the measured luminosity distance as a function of
redshift, $d_L(z) = (1+z)r(z)$. In the spatially flat universe, the comoving distance $r(z)$ from the observer to redshift $z$
is given by
\begin{equation}
\label{eq:rz} r(z;\textbf{p})=\frac{c}{H_0}\int_0^z
\frac{dz'}{E(z';\textbf{p})},
\end{equation}
wherein $\textbf{p}$ denotes the parameter space of the considered cosmological model, and $E = H/H_0$ is the dimensionless Hubble parameter.

Here, we use the ``joint light-curve analysis'' (JLA) compilation of SNe Ia \cite{Betoule2014}, which is a joint analysis of SNe Ia
observations including several low-redshift samples ($z < 0.1$), all three seasons from the SDSS-II ($0.05 < z < 0.4$), three years
from SNLS ($0.2 < z < 1$), and 14 very high redshift ($0.7 < z < 1.4$) from the Hubble Space Telescope (HST) observations.
It totals 740 spectroscopically confirmed SNe Ia with high quality light curves. In Ref~\cite{Betoule2014},
SALT2 light-curve model~\cite{Guy2007, Guy2010} have been used to fit the supernova light curves of the JLA sample.
From the observational point of view, the distance modulus of a SN Ia can be yielded from its light curve with an empirical linear
relation:
\begin{equation}
\label{eq:mu_obs} \mu^{\rm obs}_B = m_B^{\star}-(M_B-\alpha \times X_1+\beta \times \mathcal{C})
\end{equation}
The light-curve parameters $(m_B^{\star}, X_1, \mathcal{C})$ result from the fit of
the SALT2 light-curve model to the photometric data, where $m_B^{\star}$ corresponds to the observed peak magnitude in rest-frame
\emph{B} band, $X_1$ describes the time stretching of the light-curve, and $\mathcal{C}$ describes the supernova color at maximum brightness. $\alpha$, $\beta$ and $M_B$ are nuisance parameters in the distance estimate, which
are estimated simultaneously with the cosmological parameters and then marginalized over
when obtaining the parameters of interest, wherein  $M_B$ is the absolute B-band magnitude.
The theoretical (predicted) distance modulus is
\begin{equation}
\label{eq:mu_th} \mu^{\rm th}(z; \textbf{p}, \mu_0 )=5\log_{10} [D_L(z;
\textbf{p})]+\mu_0,
\end{equation}
where $\mu_0=42.38-5\log_{10}h$, which is also treated as a nuisance parameter, and the Hubble-free luminosity
distance is given by
\begin{equation}
\label{eq:DL} D_L(z;\textbf{p}) \equiv \frac{H_0}{c} d_L(z)=(1+z)\int_0^z
\frac{dz'}{E(z';\textbf{p})}.
\end{equation}
The best-fit cosmological parameters from SNe Ia data are determined by minimizing
\begin{equation}
\label{eq:chi2_SNe}
\chi^2_{\textrm{SNe}}= \sum_{i,j=1}^{740}\left[\mu_B^{\textrm{obs},i}(\alpha, \beta, M_B)-\mu^{\textrm{th},i}(z_i;\textbf{p},\mu_0)\right] Cov^{-1}_{ij} \left[\mu_B^{\textrm{obs},j}(\alpha, \beta, M_B)-\mu^{\textrm{th},j}(z_j;\textbf{p},\mu_0)\right],
\end{equation}
where $Cov$ is the covariance matrix of data vector $\vec{\mu}_B^{\textrm{obs}}$. The values of the covariance matrix $Cov$ and the SALT2
fit parameters $(m_B^{\star}, X_1, \mathcal{C})$ are available from Ref. \cite{Betoule2014}.

\subsection{Results and analysis}
In our analysis, the likelihood is assumed to be Gaussian, thus we have the total likelihood
\begin{equation}
\label{eq:LH_total}
\mathcal{L} \propto e^{-\chi_{\textrm{tot}}^2/2},
\end{equation}
where $\chi_{\textrm{tot}}^2$ is constructed as
\begin{equation}
\label{eq:chi2_total}
\chi_{\textrm{tot}}^2 = \chi^2_{\textrm{GC}} + \chi^2_{\textrm{CMB}} + \chi^2_{\textrm{SNe}},
\end{equation}
wherein $\chi^2_{\textrm{GC}}$, $\chi^2_{\textrm{CMB}}$ and $\chi^2_{\textrm{SNe}}$ are given
 by Eqs. (\ref{eq:chi2_GC}), (\ref{eq:chi2_CMB}) and (\ref{eq:chi2_SNe}), respectively, and denote the contributions from galaxy clustering,
 CMB and SNe Ia data sets described above.

We derive the posterior probability distributions of parameters with Markov Chain Monte Carlo (MCMC) exploration
using the February 2015 version of CosmoMC \cite{CosmoMc}. The parameter space of the $\Lambda$CDM model is
\begin{equation}
\label{eq:P_lcdm}
\textbf{P}_{\Lambda \textrm{CDM}} \equiv
 \{\Omega_b h^2, \Omega_c h^2, 100\theta_{MC}, \tau, {\rm{ln}}(10^{10} A_s), n_s, \Sigma m_\nu\},
\end{equation}
where $\Omega_b h^2$ and $\Omega_c h^2$, respectively, stand for the baryon and CDM densities today,
$\theta_{MC}$ is an approximation to $\theta_* = r_s(z_*)/D_A(z_*)$ (i.e., the angular size
of the sound horizon at the time of decoupling) that is used in CosmoMC and is based on fitting
formulae given in \cite{Hu_Sugiyama1996}, $\tau$ refers to the Thomson scattering optical depth due to reionization,
 $n_s$ and $A_s$ are the power-law index and amplitude of the power-law scalar primordial power spectrum of curvature perturbations,
and $\Sigma m_\nu$ is the sum of neutrino masses.
The parameter space of the $\phi$CDM model is
\begin{equation}
\label{eq:P_phicdm}
\textbf{P}_{\phi \textrm{CDM}} \equiv \{\Omega_b h^2, \Omega_c h^2, 100\theta_{MC}, \tau, {\rm{ln}}(10^{10} A_s), n_s, \Sigma m_\nu, \alpha\},
\end{equation}
which has one more parameter than that of $\Lambda$CDM model, where $\alpha$ determines the steepness of the scalar field potential in the framework of $\phi$CDM model.

\begin{figure}[t]
\centering
  \includegraphics[angle=0,width=150mm]{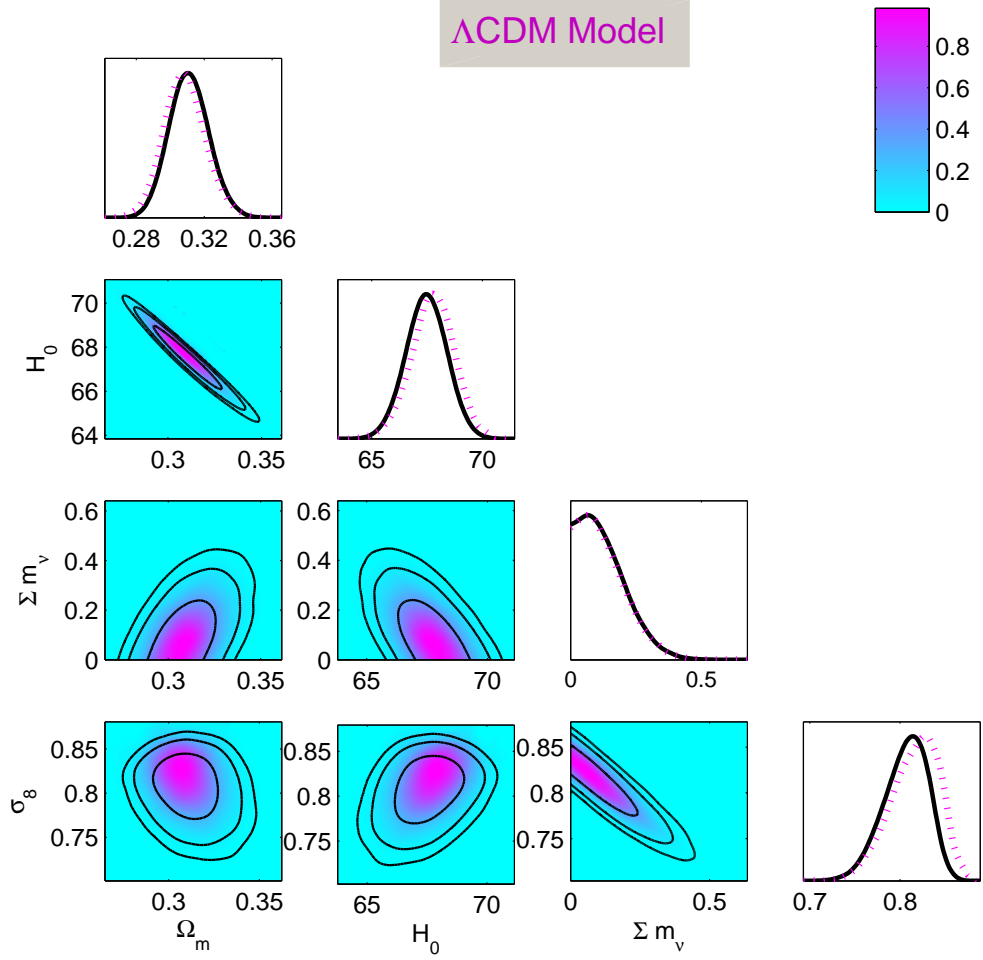}

\caption{
The 1D and 2D probability distributions of parameters of interest in the $\Lambda$CDM model constrained with
the joint sample. In the 1D plots, the solid lines denote the
 marginalized likelihoods and the dotted lines correspond to the mean likelihoods.
In the 2D plots, the contours refer to the marginalized likelihoods while the colors refer to the
 mean likelihoods. The contours correspond to 68\%, 95\% and 99\% confidence levels.
} \label{fig:LCDM_contour}
\end{figure}

\begin{figure}[t]
\centering
  \includegraphics[angle=0,width=150mm]{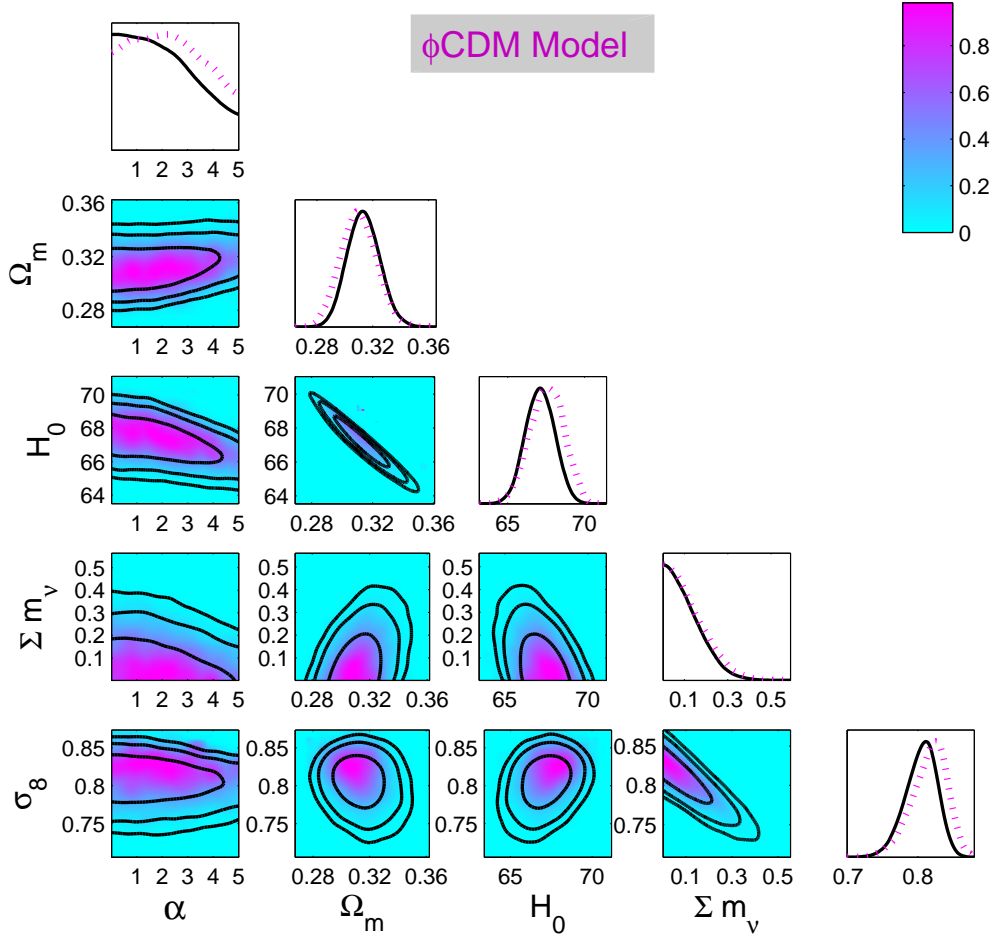}

\caption{
The 1D marginalized distribution and 2D contours of parameters of interest in the $\phi$CDM model constrained from
the joint sample. The implications of line styles and colors are the same as those in Fig. \ref{fig:LCDM_contour}.
} \label{fig:phiCDM_contour}
\end{figure}

The one-dimensional (1D) probability distributions and two-dimensional (2D) contours for the cosmological parameters of
 interest are shown in
Fig. \ref{fig:LCDM_contour} for $\Lambda$CDM model and in Fig.  \ref{fig:phiCDM_contour} for $\phi$CDM model. It shows that
constraints from the joint sample are quite restrictive, though there are degeneracies between some parameters,
such as the degeneracies in the $\Omega_m - H_0$ and  $\sigma_8 - \Sigma m_{\nu}$ planes. In addition, the differences between the
marginalized likelihoods and the mean likelihoods are modest in 1D and 2D plots. It implies that the distributions of the parameters
are almost Gaussian. We also present best-fit values and mean values with 95\% confidence limits for the parameters of interest
in Table \ref{tab:Result} both for $\Lambda$CDM and $\phi$CDM models. We find $\alpha < 4.995$ at 95\% CL for the $\phi$CDM model,
while the $\Lambda$CDM scenario corresponding to $\alpha = 0$ is not ruled out at this confidence level. The constraints on $\Omega_b h^2$, $\Omega_c h^2$, $100\theta_{MC}$,  $\tau$, ${\rm{ln}}(10^{10} A_s)$, $n_s$, $\Omega_m$, $\sigma_8$ and $H_0$ are consistent at 95\% CL for these two models.

Here, we pay attention to the constraints on $\Sigma m_\nu$. Note that $\Sigma m_\nu$ is in unit of eV.
The best-fit vale is $\Sigma m_\nu = 0.038 (0.043)$ with $\Sigma m_\nu < 0.262 (0.293)$ at 95\% CL in the framework of
$\phi$CDM ($\Lambda$CDM) model. The allowed neutrino mass scale in the $\phi$CDM model is a bit smaller than that in the $\Lambda$CDM model.
As it is shown in Figs. \ref{fig:Pk_TT_mnu} and \ref{fig:phiCDM_alpha_Pk_TT}, the increases of $\alpha$ and $\Sigma m_\nu$ both can result in
the suppression of the matter power spectrum. Therefore, when $\alpha$ is larger,
in order to avoid suppressing the matter power spectrum too much, the value of $\Sigma m_\nu$ should be smaller.
Consequently, the case with $\alpha > 0$, i.e., $\phi$CDM model, has smaller $\Sigma m_\nu$; correspondingly, the case with $\alpha = 0$,
i.e., $\Lambda$CDM scenario, has larger $\Sigma m_\nu$. Additionally, in Ref. \cite{Geng2015}, they obtained $\Sigma m_\nu < 0.200$ eV at 95\% CL in the $\Lambda$CDM model, that is consistent with our result. With the results presented in Table II of \cite{Geng2015}, one can see that our constraints on $\Omega_b h^2$, $\Omega_c h^2$,  $\tau$ and  $n_s$ are also consistent with theirs at 95\% CL for the $\Lambda$CDM model.

\section{Conclusion}
\label{summary}
We have concentrated on a quintessence model (or called as $\phi$CDM model) of dark energy with massive neutrinos.
In the $\phi$CDM model under consideration, the scalar field $\phi$ is taken as a candidate of dark energy to drive the late-time acceleration
of the universe with an inverse power-law potential $V(\phi)\propto \phi^{-\alpha}$ ($\alpha>0$).
 The larger value of $\alpha$ corresponds to the stronger time dependence of the scalar field energy density.
 When $\alpha=0$, it is reduced to the corresponding $\Lambda$CDM scenario.
The linear perturbation theory is  employed in the framework of this model. Through qualitative analyses, we find that the increases
of the sum of neutrino masses $\Sigma m_{\nu}$ and the parameter $\alpha$ both can gradually suppress the matter power spectrum $P(k)$. It implies that when the value of $\alpha$ is bigger, in order to avoid suppressing the matter power spectrum too much,
$\Sigma m_{\nu}$ should be smaller. It is in accordance with the results from the observational constraints.
The variations of these two parameters also can have signatures in the
CMB temperature anisotropy spectrum $C_l^{TT}$. In order to make a comparison, the impacts of $\Sigma m_{\nu}$
on $P(k)$ and $C_l^{TT}$ in the context of  $\Lambda$CDM model have also been presented.

A combination of the CMB data from Planck 2013 and  WMAP9, the galaxy clustering data from WiggleZ and BOSS DR11,
and the JLA compilation of the SNe Ia observations is used to constrain the parameters.
The results indicate that constraints on the cosmological parameters from this joint sample are quite restrictive.
It turns out that $\Sigma m_\nu < 0.262$ eV (95\% CL) in the framework of
$\phi$CDM  model and  $\Sigma m_\nu < 0.293$ eV (95\% CL) in the
$\Lambda$CDM model. The allowed neutrino mass scale in the $\phi$CDM model is a little shrunk comparing to that in the $\Lambda$CDM model.
In Ref. \cite{Geng2015}, it is concluded that the allowed neutrino mass scales in the viable $f(R)$ models are bigger than that
 in the $\Lambda$CDM model.
Given this, we can infer that the allowed scale of $\Sigma m_\nu$ in our $\phi$CDM model must be smaller
than those in the viable $f(R)$ models.
In addition, we get $\alpha < 4.995$ at 95\% CL for the $\phi$CDM model,  meanwhile, the $\Lambda$CDM scenario
corresponding to $\alpha = 0$ is not ruled out.
Consequently, the observational data that we have employed here still cannot distinguish whether dark energy is a time-independent
cosmological constant or a time-varying dynamical component.

\acknowledgments
Yun Chen would like to thank Bharat Ratra, Jie Liu, Chung-Chi Lee, Qiao Wang and Yuting Wang for useful discussions.
Y.C.was supported by the National Natural Science Foundation of China (Nos. 11133003 and 11573031), the Strategic Priority Research Program ``The Emergence of Cosmological Structures''
of the Chinese Academy of Sciences (No. XDB09000000), the China Postdoctoral Science Foundation (No. 2015M571126),
and the Young Researcher Grant of National Astronomical Observatories, Chinese Academy of Sciences.
L.X. was supported by the National Natural Science Foundation of China(No. 11275035), and the Open Project Program of State Key Laboratory of Theoretical Physics, Institute of Theoretical Physics, Chinese Academy of Sciences (No. Y4KF101CJ1).

\clearpage

\begin{table*}
\begin{center}{\large
\begin{tabular}{|p{3cm}|p{3cm}|p{3cm}|p{3cm}|p{3cm}|}
\hline

  \multirow{2}{2cm}{Parameters}& \multicolumn{2}{c|}{$\Lambda$CDM Model} & \multicolumn{2}{c|}{$\phi$CDM Model} \\
 \cline{2-5}
 & Best-fit values & 95\% limits & Best-fit values & 95\% limits\\

\hline
$\Omega_b h^2$ & 0.0221& $ 0.0221 \pm 0.0005$ &  0.0222 & $0.0222\pm 0.0005$\\
$\Omega_c h^2$ & 0.1197 &  $0.1180^{+0.0036}_{-0.0037}$ &  0.1176 &  $0.1177 \pm 0.0036$\\
$100\theta_{MC}$ & 1.0412 &  $1.0414 \pm 0.0011$ & 1.0414 & $1.0415^{+0.0012}_{-0.0011}$\\
$\tau$          & 0.0921 &  $0.0904^{+0.0263}_{-0.0255}$  & 0.0846 &  $0.0914^{+0.0266}_{-0.0242}$\\
${\rm{ln}}(10^{10} A_s)$ &3.0939 & $3.0854^{+0.0513}_{-0.0467}$  &  3.0758 &  $3.0869^{+0.0516}_{-0.0475}$ \\
$n_s$ & 0.9601 &  $0.9636^{+0.0113}_{-0.0112}$ &  0.9607 &  $0.9644^{+0.0118}_{-0.0119}$  \\
$\Sigma m_\nu$ &  0.043 &  $< 0.293$ &  0.038 &  $< 0.262$ \\
$\alpha$ & ... & ... & 2.482 &  $ < 4.995$  \\
\hline
$\Omega_m$ &  0.312 &  $0.311^{+0.023}_{-0.022}$ &  0.310 &  $0.313^{+0.023}_{-0.021}$ \\
$\sigma_8$ & 0.834 &  $0.806^{+0.044}_{-0.049}$ & 0.806 &  $0.805^{+0.043}_{-0.046}$  \\
$H_0$ &  67.51 &  $ 67.47^{+1.76}_{-1.79}$ &  67.33 &  $ 67.11^{+1.79}_{-1.81}$ \\

\hline

\end{tabular}}\\
\end{center}
\caption{Fitting results from the joint sample.  We present the best-fit values (i.e., the parameters that maximize the overall
likelihood), and the mean values with 95\% confidence limits for the parameters of interest. Where $\Sigma m_\nu$ is
in unit of eV, and $H_0$ is in unit of km/s/Mpc. The top block contains parameters with uniform priors that are
varied in the MCMC chains. The lower block defines various
derived parameters.}\label{tab:Result}
\end{table*}

\end{document}